\documentclass{iopart}
\usepackage{iopams} 
\usepackage{epsfig}  
\usepackage{graphics}  

\newcommand{\Ai}{\rm{Ai}}
\newcommand{\AiInt}{\rm{Ai}_1}
\newcommand{\He}{\rm{H}}
\newcommand{\La}{\rm{L}}
\newcommand{\erfc}{\rm{erfc}}

\newcommand{\mbfo}{\mathbf{o}}

\newcommand{\mbfH}{\mathbf{H}}

\newcommand{\elf}{{\cal E}_\perp}
\newcommand{\ELF}{\boldsymbol{\cal E}}
\newcommand{\mgf}{{\cal B}}
\newcommand{\MGF}{\boldsymbol{\cal B}}

\begin{document}

\title{Electron propagation in crossed magnetic and electric fields}

\author{T. Kramer\dag, C. Bracher\ddag, and M. Kleber\dag}

\address{\dag\ Physik-Department T30,
Technische Universit\"at M\"unchen,
James-Franck-Stra{\ss}e,
85747 Garching, Germany}
\address{\ddag\ Department of Physics and Atmospheric Science, Dalhousie University, Halifax, N.S.\ B3H~3J5, Canada}
\ead{tkramer@ph.tum.de}

\begin{abstract}
Laser--atom interaction can be an efficient mechanism for the production of coherent electrons. We analyze the dynamics of monoenergetic  electrons in the presence of uniform, perpendicular magnetic and electric fields. The Green function technique is used to derive analytic results for the field--induced quantum mechanical drift motion of i) single electrons and ii) a dilute Fermi gas of electrons. The method yields the drift current and, at the same time it allows us to quantitatively establish the broadening of the (magnetic) Landau levels due to the electric field: Level number $k$ is split into $k+1$ sublevels that render the $k$th oscillator eigenstate in energy space.  Adjacent Landau levels will overlap if the electric field exceeds a critical strength. Our observations are relevant for quantum Hall configurations whenever electric field effects should be taken into account. 
\end{abstract}

\pacs{73.20.At,03.75.-b,73.43.Cd}

\section{Introduction}

The quasi--free propagation of electrons in strong external laser fields is known to play an important role for the interpretation of phenomena that cannot be described in perturbation theory: The quantum propagation of photoelectrons in a laser field is essential for the interpretation of non--linear effects such as the generation of plateaus in high--harmonic generation \cite{Agostini1979a,Corkum1993a,Becker1995a}. More generally, one expects that a study of the quantum motion of matter waves in external fields will shed light on the occurrence of interesting interference phenomena. In addition, external fields offer a useful testing ground for analyzing the interplay between classical and quantum dynamics.

An impressive example for the application of a static external electric field in basic laser--atom physics is the so--called photodetachment microscope \cite{Blondel1996a} which is used to determine electron affinities of negative ions with high accuracy.  Similarly photodetachment studies in a magnetic field reveal oscillations in the photocurrent spectrum as a function of the energy of the emitted photoelectrons (for a collection of references and new experiments see \cite{Yukich2003a}). The observed oscillatory structure is clearly related to the existence of Landau levels. In this paper we will analyze the quantum motion of electrons in crossed static magnetic and electric fields. One expects the subtle properties of a magnetic field to show up most clearly in two-dimensional motion in a plane perpendicular to the magnetic field. In this way one generates a Hall configuration with the electric field $\ELF$ (Hall field) orthogonal to the magnetic field $\MGF$.

A convenient way of achieving this goal consists in utilizing the method of Green functions. The Green function $G(\mathbf r,\mathbf r';E)$ is linked to the familiar time--dependent quantum propagator $K(\mathbf{r},t|\mathbf{r'},0)$
\cite{Feynman1965a} by means of a Laplace transform \cite{Grosche1998a}:
\begin{equation}
\label{eq:defGr}
G(\mathbf{r},\mathbf{r'};E)=
\frac{1}{{\rm i}\hbar}
\int_0^\infty {\rm d}t\,
{\rm e}^{{\rm i}E t/\hbar}
K(\mathbf{r},t|\mathbf{r'},0),
\end{equation}
where the propagator represents the probability amplitude for a particle to travel from $\mathbf{r}$ to $\mathbf{r'}$ in a fixed time $t$. The energy Green function corresponds to the same quantum travel, however with fixed energy $E$. It is important to realize that (\ref{eq:defGr}) will automatically lead to a retarded Green function, with a matter wave being generated at $\mathbf{r}'$ with part of it subsequently travelling to $\mathbf{r}$. For \emph{free} particles the Green function is the well--known outgoing wave,
\begin{equation}
\label{eq:wave}
G(\mathbf{r},\mathbf{r'};E) = - \frac m{2\pi\hbar^2}
\frac{\exp(ik|\mathbf{r} -\mathbf{r'}|)}{|\mathbf{r} -\mathbf{r'}|}
\end{equation}
with $k=\sqrt{2mE}/\hbar$ being the wave number. In external fields, (\ref{eq:wave}) will be modified as the wave leaves the source point $\mathbf{r'}$. If there are $K$ classical ($cl$) trajectories connecting $\mathbf{r'}$ and $\mathbf{r}$, we have to sum over $K$ amplitudes in order to obtain the Green function in semiclassical ($sc$) approximation:
\begin{equation}
\label{eq:semic}
G_{sc}(\mathbf{r},\mathbf{r'};E)=
\sum_{k=0}^{K} A_k\exp\left(\frac{i}{\hbar}W_{cl}^{k}(\mathbf{r},\mathbf{r'};E)\right).
\end{equation}

For the quadratic potentials considered here, the amplitudes $A_k$ are independent of $\mathbf{r}$ and $\mathbf{r'}$ and are not written down explicitly. What matters here is Hamilton's characteristic function $W_{cl}^{k}(\mathbf{r},\mathbf{r'};E)$ for path number $k$. $W$ is also known as the reduced classical action. It is related to the classical action $S_{cl}(\mathbf{r},\mathbf{r'};t_k)$ via the Legendre transformation:
\begin{equation}
\label{eq:actio}
S_{cl}(\mathbf{r},\mathbf{r'};t_k)=
W_{cl}^{k}(\mathbf{r},\mathbf{r'};E) - Et_k .
\end{equation}
Here $t_k$ is the travel time along the classical path number $k$. Since the Green function can be interpreted as the wave function of the moving electron, with $\mathbf{r'}$ being the starting point, we can readily calculate the current density $\mathbf{j}_{sc}(\mathbf{r})$ of the electron. 
According to the laws of classical mechanics, the classical momentum of a particle can be derived from Hamilton's characteristic function:
\begin{equation}
\label{eq:pulse}
\mathbf{p}_{cl}^{k}=
\mathbf{\nabla}_{\mathbf{r}}W_{cl}^{k}(\mathbf{r},\mathbf{r'};E).
\end{equation}
It is well known that cycloids describe the classical motion of classical electrons in a combined electric and magnetic field.
However, the \,\emph{average} classical velocity of an electron moving in a plane orthogonal to the $\MGF$--field is the drift velocity, $|\mathbf{v}_{d}| = |\ELF|/|\MGF|$, which is the same for each path $k$. In view of (\ref{eq:semic}) we may therefore write
\begin{equation}
\label{eq:strom}
\langle \mathbf{j}_{cl}(\mathbf{r})\rangle =
e \,\langle\rho(\mathbf{r})\rangle \mathbf{v}_{d} ,
\end{equation}
where $\mathbf{j}$ and $\rho(\mathbf{r}) =  {|G_{sc}(\mathbf{r},\mathbf{r'};E)|}^2$ have been properly averaged. 

What we have demonstrated here is that, by calculating the Green function with outgoing--wave boundary conditions we will automatically obtain the $\ELF\times\MGF$--drift in the classical limit. As will be shown below, the Green function also contains information to what extent an electron with energy $E$ will participate in the drift motion. To put it in a more formal way: By utilizing the appropriate Green function we can calculate the local density of states $n(\mathbf{r},E)$, which plays a central role for calculating the spatial flow pattern of matter waves in external fields. To see this more clearly, we rewrite the operator equation for $G$,
\begin{equation}
\label{eq:Gdiff}
(E-H)\,G(\mathbf{r},\mathbf{r'};E)= \delta(\mathbf{r} -\mathbf{r'})
\end{equation}
in terms of a formal solution
\begin{equation}
\label{eq:Gform}
G(\mathbf{r},\mathbf{r'};E)= \, \left\langle \mathbf{r} \left| \frac{1}{E - H + i\epsilon} \right| \mathbf{r'} \right\rangle \,
\end{equation}
where $\epsilon = 0^+$ takes care of the outgoing boundary condition. An eigenstate expansion of (\ref{eq:Gform}) then yields 
\begin{equation}
\label{eq:Geign}
G(\mathbf{r},\mathbf{r'};E)= 
\sum_{l}\frac{\langle\mathbf{r}|\Psi_{l}\rangle \langle\Psi_{l}|\mathbf{r'}\rangle}{E - E_{l} + i\epsilon} \,
\end{equation}
where $\Psi_{l}(\mathbf{r}) =\langle\mathbf{r}|\Psi_{l}\rangle$ is the $l$th eigenfunction of H with energy $E_{l}$. Defining the so--called local density of states
\begin{equation}
n(\mathbf{r},E)\, := 
\sum_{l}|\Psi_{l}(\mathbf{r})|^2\,\delta(E-E_l)
\end{equation}
we obtain from (\ref{eq:Gform})
\begin{equation}
\label{eq:dense}
n(\mathbf{r},E)\, = 
-\frac{1}{\pi}\,\mbox{Im}\{G(\mathbf{r},\mathbf{r};E )\}
\end{equation}
It is the imaginary part of the Green function which governs the current of the corresponding quantum motion. Therefore, the magnitude of $n(\mathbf{r},E)$ determines to which extent an electron that travels through $\mathbf{r}$ with an energy $E$ will participate in the motion (here the $\ELF\times\MGF$ drift).

From the gauge property of the Green function in static electromagnetic fields,
\begin{equation}
\label{eq:gauge}
G_{\ELF,\MGF}(\mathbf{r},\mathbf{r'};E)= \,\exp\left[\frac{iq}{2\hbar c} \MGF\cdot (\mathbf{r'}\times \mathbf{r}) \right] G_{\ELF,\MGF}(\mathbf{r}- \mathbf{r'},\mathbf{o};E + q \mathbf{r'}\cdot\ELF) \,
\end{equation} 
with $q=-e$ in our case.
In the following we will call 
\begin{equation}
\label{eq:short}
n(E) =  n(\mathbf r = \mathbf o,E)\,
\end{equation}
the density of states because $n(\mathbf{r},E)$ is readily obtained from (\ref{eq:gauge}).
In this paper we will calculate the density of states $n(E)$ for two-dimensional motion in perpendicular, homogeneous electric and magnetic fields. All necessary results will be derived here; they are based on a more general approach that is described in detail in \cite{Kramer2001a,Bracher2003a}. For a wave function approach we would like to refer to \cite{Johnson1983a}.

\section{Electric field effects in two--dimensional motion}
In the absence of an electric field, the three-dimensional Green function for a uniform magnetic field was obtained in closed form in \cite{Dodonov1975a}. Also, for a purely magnetic field, the two-dimensional density of states  is known \cite{Prange1987a,Grosso2000a} to show a spike-like structure, formally written as a superposition of discrete $\delta$--distributions positioned at the Landau levels at $E=(2k+1)\hbar\omega_L$, where $\omega_L=e\mgf /(2m)$ denotes the Larmor frequency:
\begin{equation}\label{eq:DOSB}
n_{\MGF}^{(2D)}(E)=\frac{e \mgf }{2\pi\hbar}\sum_{k=0}^\infty
\delta\left(E-\hbar\omega_L[2k+1]\right).
\end{equation}
As we will show below, the presence of an electric field broadens the $\delta$--distribution of the $k$th Landau level into a smooth function of overall Gaussian form that is modulated into $k+1$ distinct sublevels.

The desired propagator is governed by the Hamiltonian of a 2D-electron in crossed fields,
\begin{equation}\label{eq:HEB}
\fl \mbfH_{\ELF\times\MGF}^{(2D)} = \frac{p_x^2+p_y^2}{2m} +\frac{1}{2}m\omega_L^2\left(x^2+y^2\right) - \mathbf r_\perp\cdot\mathbf F_\perp - p_y x \omega_L + p_x y \omega_L,
\end{equation}
where $\mathbf F_\perp = -e \boldsymbol\elf$ denotes the electric force in the $x$--$y$-plane. The result is \cite{Nieto1992a}:
\begin{equation}\label{eq:JCrossedInt2D}
\fl K_{\ELF\times\MGF}^{(2D)}(\mbfo,t|\mbfo,0) = -\frac{\rmi m\omega_L}{2\pi\hbar\sin(\omega_L t)}
\exp \left\{ \frac{\rmi F_\perp^2 t}{8m\hbar\omega_L^2}
\left[ \omega_Lt \cot\left(\omega_{L}t\right) - 1 \right]
\right\} .
\end{equation}
To account for the effects of the electron spin, we note that the spin adds a constant term $\pm \frac{1}{2}g \mu_B\mgf = \pm \frac{1}{2}g\hbar\omega_L$ to the Hamiltonian and thus merely shifts the energy by this amount \cite{Prange1987a}. The spin dependent densities of states become
\begin{equation}
\label{eq:spin}
n_{\uparrow,\downarrow}(E) =n\left(E \pm \frac{1}{2}g\hbar\omega_L\right)
\end{equation}
and the total density of states including spin can be mapped back to the one without spin: $n_{\uparrow\downarrow}(E)=n_\uparrow(E)+n_\downarrow(E)$.  Thus we evaluate $n(E)$ and defer the inclusion of spin for the moment.

For the discussion of the density of states in crossed fields it is useful to transform the trigonometric functions in the propagator into a sum over Landau levels. This can be done using the identity
\begin{equation}
\frac{\exp[-\alpha\coth(z)]}{\sinh(z)} = 2\rme^{-\alpha} \sum_{k=0}^\infty
\La_k^{(0)}(2\alpha)\exp\left[-2z \left(k+ \frac12 \right)\right],
\end{equation}
which follows after substituting $t=\exp(-z)$ in the generating function of the Laguerre polynomials $\La_k^{(0)}(z)$ \cite{Abramowitz1965a}.  The imaginary part of the Green function therefore reads
\begin{equation}
\label{eq:DOSEBa}
n_{\ELF\times\MGF}(E) = -\frac{1}{\pi} \Im\left[G_{\ELF\times\MGF}^{(2D)}(\mbfo,\mbfo;E)\right] =
\sum_{k=0}^\infty n_{k,\ELF\times\MGF}(E),
\end{equation}
where the partial density of states assigned to the $k$th Landau level is given by:
\begin{eqnarray}
\label{eq:DOSEB}
n_{k,\ELF\times\MGF}(E) &=&
\frac{m\omega_L}{2\pi^2\hbar^2}
\int_{-\infty}^\infty\rmd t\; \rme^{-\Gamma^2 t^2/(4\hbar^2)-\rmi t E_k/\hbar}\, \La_k^{(0)} \left(\frac{\Gamma^2 t^2}{2\hbar^2}\right)\nonumber\\
&=& \frac1{2^{k+1} k! \pi^{3/2}l^2\Gamma} \, \rme^{-E_k^2/\Gamma^2} \, {\left[\He_k\left(E_k/\Gamma\right)\right]}^2 .
\end{eqnarray}
Here, the level width parameter
\begin{equation}
\Gamma = F_\perp l
\end{equation}
is related to the magnetic length $l = \sqrt{\hbar/(e\mgf)}$, and $E_k$ denotes the effective energy shift for the $k$th level:
\begin{equation}
E_k=E-\Gamma^2/(4\hbar\omega_L)-(2k+1)\hbar\omega_L.
\end{equation}
In eq.~(\ref{eq:DOSEB}), we evaluated the integral over $t$ and expressed the result in terms of Hermite polynomials $\He_k(z)$ \cite{Abramowitz1965a}. The density of states is identical to a sum over the probability densities of the eigenstates of a one-dimensional harmonic oscillator
\begin{equation}\label{eq:QHO}
{\left|u_k(\xi)\right|}^2 = \frac1{2^k k!\,\sqrt\pi} \,
\rme^{-\xi^2}
{\left[ \He_k\left(\xi\right)\right]}^2 .
\end{equation}
It should be emphasized that the oscillator functions depend here on the energy $E$ and not on a space variable.  The dimensionless energy variable reads $\xi=E_k/\Gamma=E_k/(F_\perp l)$,  with the $k$th eigenstate being centered around the $k$th Landau level, apart from an overall shift $\Gamma^2/(4\hbar\omega_L) = \frac m2 (\elf/\mgf )^2$ accounting for the kinetic energy of the drifting electrons. This result is also consistent with the mapping of the original Hamiltonian in crossed fields in eq.~(\ref{eq:HEB}) to the Hamiltonian of a shifted harmonic oscillator. Details of the corresponding canonical transformation and its unitary representation are presented in Ref.~\cite{Kramer2000a}.  The total contribution of the $k$th Landau level integrated over energy-space is readily available from the normalization of the oscillator eigenstates:
\begin{equation}\label{eq:EBQ}
\int_{-\infty}^\infty\rmd E\;
n_{k,\ELF\times\MGF}(E)=
\frac{e\mgf }{2\pi\hbar}
\int_{-\infty}^\infty\rmd \xi\;{\left|u_k(\xi)\right|}^2 = \frac{e\mgf }{2\pi\hbar}.
\end{equation}
This result reflects the quantization of each Landau level in a purely magnetic field, as given by eq.~(\ref{eq:DOSB}).  In the context with fermionic matter waves, as discussed in the next section,  there is another important quantity: the energy--integrated density of states $N(E_F)$, integrated up to the Fermi energy $E_F$ of the system,
\begin{equation}\label{eq:IDOS}
N(E_F)=\int_{-\infty}^{E_F}\rmd E\;n(E).
\end{equation}
(A recursive evaluation scheme allows to express $N(E_F)$ as a sum over error functions, but we omit the result here.)

\begin{figure}
\includegraphics[width=0.5\textwidth]{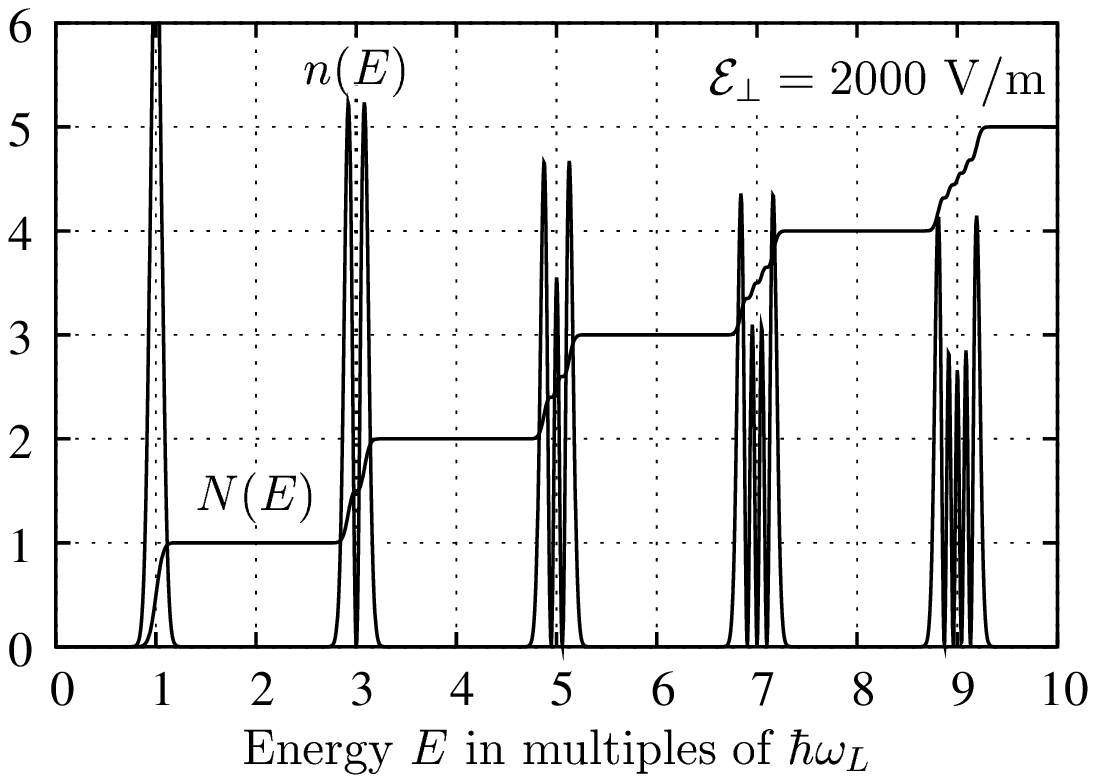}
\includegraphics[width=0.5\textwidth]{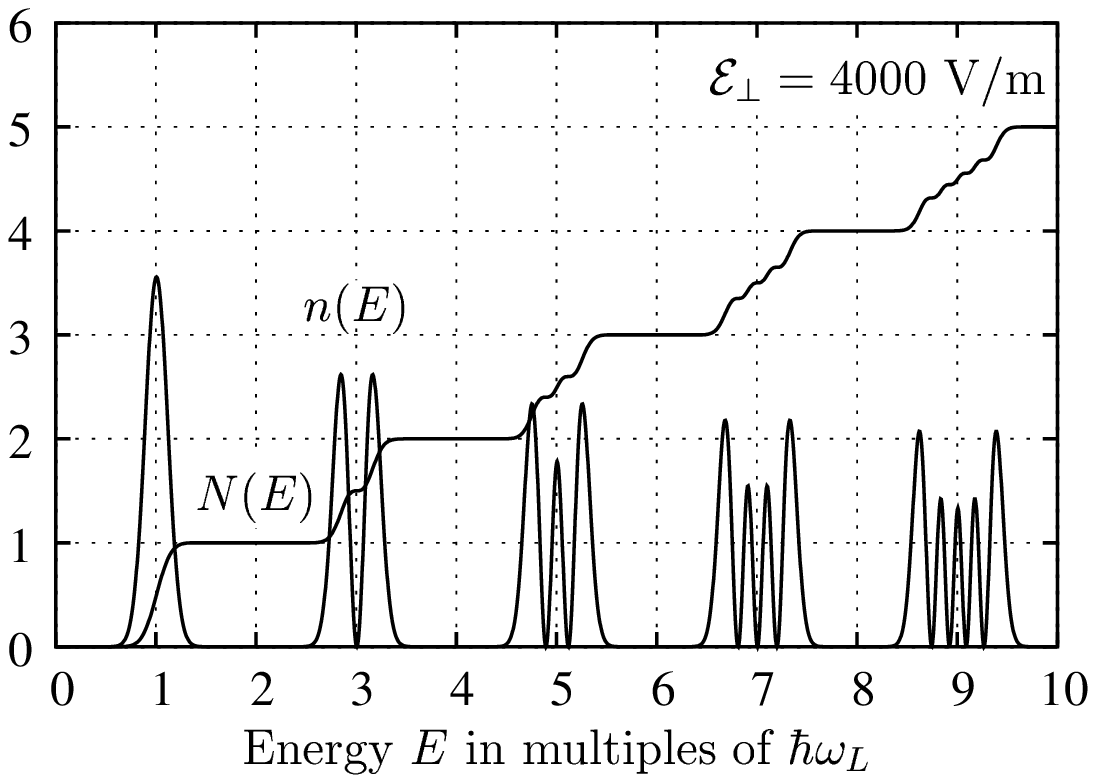}\\
\includegraphics[width=0.5\textwidth]{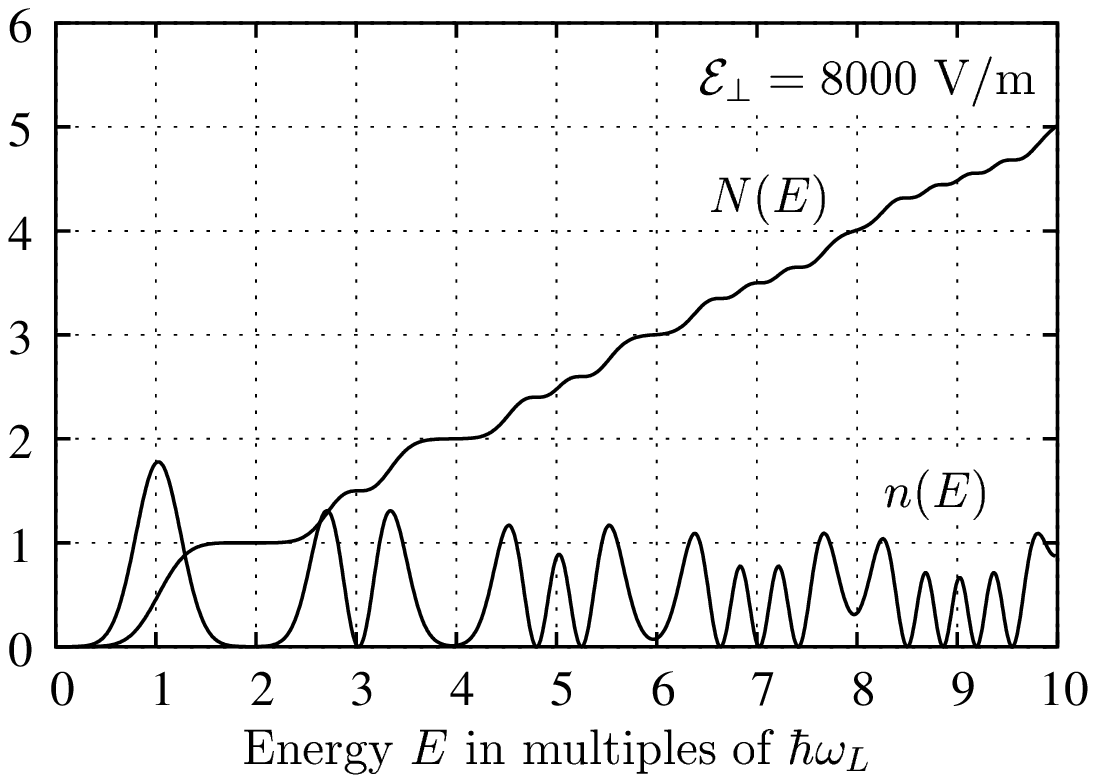}
\includegraphics[width=0.5\textwidth]{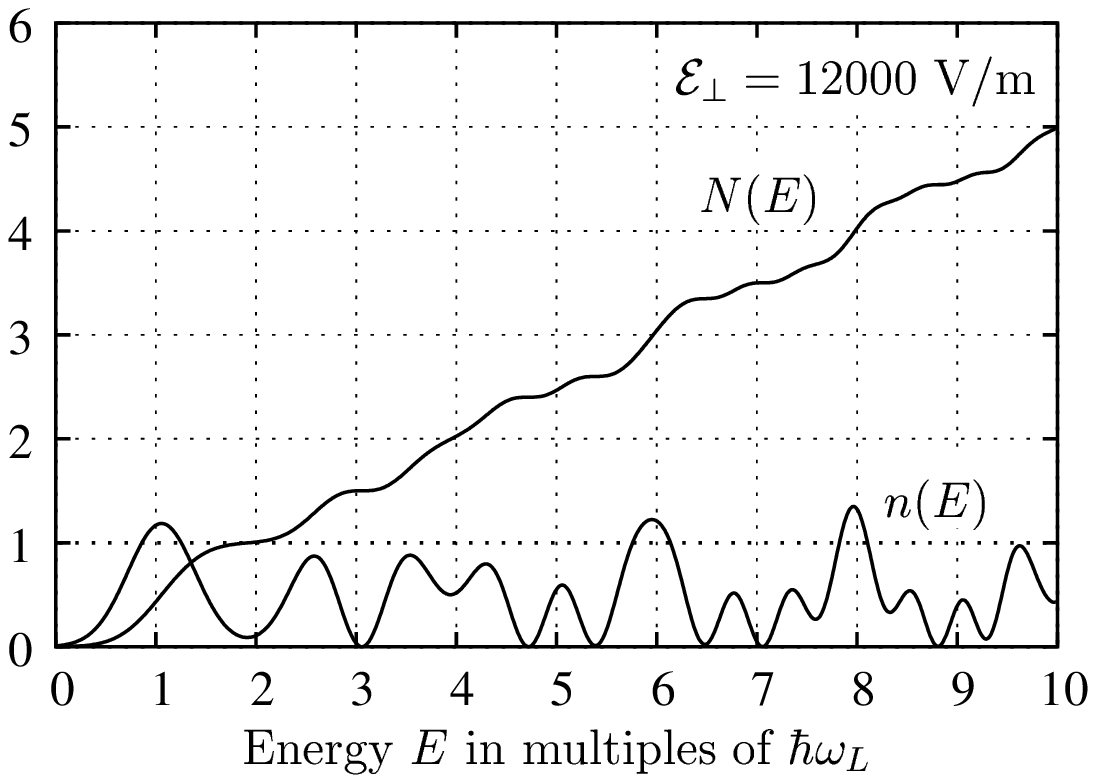}
\caption{Two-dimensional density of states $n(E)$ (in units of $e\mgf /(2\pi\hbar^2\omega_L)$) and integrated density of states $N(E)$ (in units of $e\mgf /(2\pi\hbar)$) at four different electric fields $\elf=2000, 4000, 8000,12000$~V/m and for a magnetic field $\mgf =5$~T as a function of the scaled energy $E/(\hbar\omega_L)$ according to eq.~(\ref{eq:DOSEB}). Near the $k$th Landau level at $E=(2k+1)\hbar\omega_L$, the density of states renders the probability distribution of a one-dimensional harmonic oscillator in the $k$th eigenstate.}
\label{fig:DOS}
\end{figure}

For each Landau level $k$, the density of states has a Gaussian envelope with width $\Gamma$ that is split into $k+1$ intervals by the $k$ simple zeroes $\xi_{k,j}$ ($j=1,\ldots,k$) of the polynomial $\He_k(\xi)$.  In fig.~\ref{fig:DOS}, we plot the resulting density of states for various electric field strengths $\elf$.  For small $\elf$, the overlap between adjacent Landau levels is negligible, as the density of states drops off exponentially between them.  With increasing electric field, the Landau levels broaden and finally coalesce.  
We infer from eq.~(\ref{eq:QHO}) that the classical turning point, $\xi^{\rm tp}_k=\sqrt{2k+1}$, provides a practical measure for the width of the partial density of states $n_{k,\ELF\times\MGF}(E)$. The classically allowed region in energy between two Landau levels $k-1$, $k$ is then given by the ratio:
\begin{equation}\label{eq:LLoverlap}
\fl
\frac{{\rm half~widths~of~adjacent~levels}\;\Gamma(\xi^{\rm tp}_{k-1}+\xi^{\rm tp}_k)}{{\rm level~spacing}\;2\hbar\omega_L}
= \frac{m}{\sqrt{e\hbar}} \left(\sqrt{2k-1}+\sqrt{2k+1}\right) \frac{\elf}{\mgf^{3/2}}.
\end{equation}
The overall width of the modulated Landau levels increases with $k^{1/2}$.  Note that all features of the level density of states, including the nodes, scale linearly in width with the electric field $\elf$.
We conclude that for small magnetic fields, many Landau levels overlap, and the density of states becomes smooth.  In the limit $\mgf \rightarrow 0$, an exact expression for the two-dimensional density of states in a purely electric field may be extracted from eq.~(\ref{eq:JCrossedInt2D}):
\begin{equation}
\label{eq:DOS2Delectric}
n_{\ELF}(E) = -\frac1\pi \Im\left[G_{\ELF}^{(2D)}(\mbfo,\mbfo;E)\right]
= \frac m{2\pi\hbar^2} \left[ \frac13 - \AiInt\left( - \Lambda E \right)  \right],
\end{equation}
where $\Lambda = 2 [m/(\hbar e \elf)^2]^{1/3}$, and $\AiInt(z) = \int_0^z \rmd x\, \Ai(x)$ denotes the integral of the Airy function \cite{Abramowitz1965a, Bracher2003a}.
If additionally $\elf\rightarrow 0$, the constant density of states of a free two-dimensional electron gas is recovered: $n_{\rm free}(E) = \Theta(E)\cdot m/(2\pi\hbar^2)$.

\section{Quantum motion of a dilute Fermi gas in two dimensions}

For a non--interacting, i.e.\ dilute, gas of charged fermions that move against a uniformly charged background, the Pauli principle prevents more than single occupancy of each energy quantum state. For such a dilute low--temperature gas, the available states are filled up to the Fermi level. Hence, the number of occupied states is given by (\ref{eq:IDOS}). Since the electrons are flowing at a right angle to the electric field and not parallel to $\ELF_\perp$, it is possible to introduce a local Fermi level \cite{Halperin1986a}
\begin{equation}
\label{eq:tilt}
E_F(\mathbf{r})= E_F(\mathbf{o}) - q \mathbf{r}\cdot\ELF
\end{equation}
which takes into account that all energy levels are tilted in the direction of the electric field. In view of (\ref{eq:gauge}) we conclude that there is translational invariance in the sense that 
\begin{equation}
\label{eq:tilt1}
G_{\ELF\times\MGF}(\mathbf{r},\mathbf{r};E_{F}(\mathbf{r}))= \,
G_{\ELF\times\MGF}(\mathbf{o},\mathbf{o};E_{F}(\mathbf{o})) \,.
\end{equation} 

The exponential suppression of the density of states between Landau levels has profound implications on the conductivity in a two-dimensional system, including the resistivity plateaus observed in the quantum Hall effect (QHE), as we show now.  In a simple Drude-like model we may model the dynamics of the electrons by the electric and Lorentz forces amended with a term that incorporates elastic scattering via a relaxation time $\tau(E)$:
\begin{equation}
m\frac{\rmd \mathbf{v}}{\rmd t}=e\ELF+e\mathbf{v}\times\mgf-\frac{m}{\tau}\mathbf{v}.
\end{equation}
Under stationary conditions $\mathbf{v}$ is constant and together with the current density $\mathbf{j}=ne\mathbf{v}$ the components of the conductivity tensor $\mathbf{j}=\sigma\ELF$ become \cite{Grosso2000a}
\begin{equation}
\sigma(E)=\frac{e^2 n(E)\tau(E)}{m}
\frac{1}{1+\omega_C^2\tau(E)^2}
\left(
\begin{array}{cc}
1&-\omega_C\tau(E) \\ \omega_C\tau(E)&1
\end{array}
\right),
\end{equation}
where $\omega_C=2\omega_L=eB/m$. For $T\rightarrow 0$, the total conductivity is obtained by integrating over the occupied energy range
\begin{equation}
\sigma=\int_{-\infty}^{E_F}\rmd E\;\sigma(E).
\end{equation}
The availability of empty states limits significant scattering to an energy range of order $k_B T$ around $E_F$.  For energies $E$ far from the Fermi energy, $\tau(E)\rightarrow\infty$.  In strong magnetic fields, the relaxation time $\tau(E)$ thus satisfies ${[\omega_C\tau(E)]}^2 \gg 1$ and the transversal component $\sigma_{xy}$ mirrors the integrated density of states $N(E_F)$
\begin{equation}
\sigma_{xy}=
\frac{e}{\mgf }\int_{-\infty}^{E_F}\rmd E\;
\frac{n(E)}{1+[\omega_C\tau(E)]^{-2}}
=\frac{e}{\mgf }N(E_F),
\end{equation}
whereas the longitudinal conductivity $\sigma_{xx}$ is dominated by contributions in the vicinity of the Fermi energy:
\begin{equation}
\sigma_{xx} = \frac{e}{\mgf }\int_{-\infty}^{E_F} \rmd E\;
\frac{n(E) \omega_C\tau(E)}{1+\omega_C^2\tau(E)^2}
\approx k_B T \frac{e}{\mgf} n(E_F) 
\frac{\omega_C\tau(E_F)}{1+\omega_C^2\tau(E_F)^2}.
\end{equation}
Thus, the longitudinal conductivity is proportional to the density of states at $E_F$.  Finally, we note that the resistivity tensor is related to $\sigma_{xx}$ and $\sigma_{xy}$ via:
\begin{equation}
\rho_{xy}=\frac{\sigma_{xy}}{\sigma_{xx}^2+\sigma_{xy}^2},\qquad
\rho_{xx}=\frac{\sigma_{xx}}{\sigma_{xx}^2+\sigma_{xy}^2}.
\end{equation}
From eq.~(\ref{eq:EBQ}), it is clear that $\rho_{xy}$ is quantized between Landau levels.  If $E_F$ lies in the gap between the $(k-1)$th and $k$th level, $n(E_F)\approx0$ is exponentially suppressed, $N(E_F)=ke\mgf /(2\pi\hbar)$ and therefore
\begin{equation}
\label{eq:integerQHE}
\rho_{xx} \approx 0 , \qquad \rho_{xy} \approx \frac{2\pi\hbar}{k e^2},
\end{equation}
where, at $T\rightarrow 0$, the deviations are of the order $\exp(-\hbar^2\omega_L^2/\Gamma^2) = \exp[-e\hbar \mgf ^3 /(4m^2\elf^2)]$.
\begin{figure}
\includegraphics[width=0.897\textwidth]{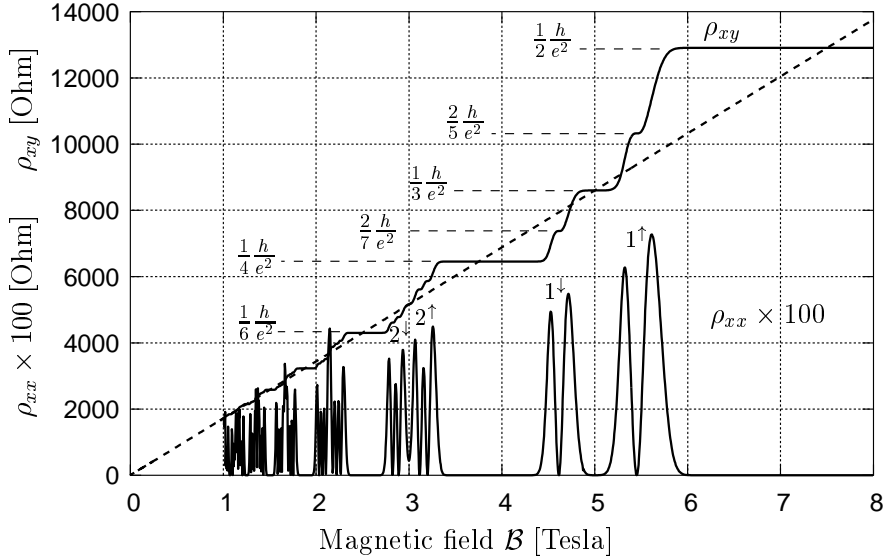}
\caption{Quantum Hall effect at strong magnetic fields ($B>1$~Tesla) for a non-interacting two-dimensional electron gas.  Hall resistance $\rho_{xy}$ and longitudinal resistance $\rho_{xx}$ as a function of the magnetic field $\mgf $ for fixed Fermi energy ($E_F=0.868$~meV).  Effective mass $m^*=1$, effective $g$-factor $g=\frac{1}{2}$, current density $j_x=0.2$~A/m, $\tau(E_F)=10^{-11}$~s, $T=0.1$~K. The dashed line represents the classical Hall resistance $\rho_{xy}$ with a constant level density. Similar experimental results are shown i.e.\ in \cite{Paalanen1982a}. However, we assumed a high current density in order to show the splitting of the Landau levels clearly.}
\label{fig:HallB}
\end{figure}
It is interesting to see how single--particle effects can contribute to the existence of conductivity quantization.
\subsection{Non-integer filling factors}

We now exploit the fact that the density of states does not only vanish between Landau levels, but also within these levels whenever $E_F$ coincides with one of the zeroes $\xi_{k,j}$ of the individual Hermite polynomials $\He_k\left(E_k/\Gamma\right)$ in eq.~(\ref{eq:DOSEB}).  In the vicinity of these zeroes, an additional plateau-like structure in the Hall resistivity $\rho_{xy}$ will appear. There, the suppression of the density of states is only quadratic in $E$. Therefore, these plateaus behave differently from the ones due to the integer quantum Hall effect in eq.~(\ref{eq:integerQHE}).  Continuing the notation employed above, the new set of plateaus is characterized by non-integer filling factors $\kappa$:
\begin{equation}
\label{eq:non-integerQHE}
\rho_{xy} = \frac{2\pi\hbar}{\kappa_{k,j} e^2}.
\end{equation}
The values of $\kappa$ are obtained by integrating the density of states up to the zeroes of the Hermite polynomial. A non-integer $\kappa$ is closely related to the probability of finding a particle between the nodes of the $k$th oscillator eigenstate in eq.~(\ref{eq:QHO}).  Adding $\xi_{k,0}=-\infty$, $\xi_{k,k+1}=\infty$ to the list of zeroes of $u_k(\xi)$, the contribution $\Delta_{k,j}$ of the interval between two neighbouring nodes in the $k$th level becomes
\begin{equation}
\label{eq:non-integer1}
\Delta_{k,j} = \int_{\xi_{k,j-1}}^{\xi_{k,j}} \rmd \xi\; {\left|u_k(\xi)\right|}^2,
\quad{\rm yielding}\quad
\kappa_{k,j}=k+\sum_{i=1}^{j}\Delta_{k,i}\,.
\end{equation}
Note that the sum $\sum_{i=1}^{k+1}\Delta_{k,i} = 1$ represents the entire level $k$ and thus must attain unit value.  For $j=0$, $\kappa_{k,0} = k$, and the integer quantum Hall effect in eq.~(\ref{eq:integerQHE}) is recovered, while the condition $j>0$ yields fractional values for $\kappa_{k,j}$.  The parity $(-1)^k$ of the functions $u_k(\xi)$ implies $u_k(0)=0$ for odd $k$, and thus the existence of half-integer filling factors:  $\kappa_{1,1}=3/2$, $\kappa_{3,2}=7/2$, etc.  The other intervals lead to irrational filling factors that, however, are surprisingly close to simple fractions. They occur for $k\geq2$.  In the third Landau level ($k=2$), we obtain with $\xi_{2,2} = -\xi_{2,1} = 1/\sqrt 2$ from eq.~(\ref{eq:non-integer1}):
\begin{equation}
\label{eq:non-integer2}
\Delta_{2,1} = \Delta_{2,3} = \frac1{\sqrt{2\pi\rme}} + \frac12 \erfc\left( \frac1{\sqrt2} \right) = 0.400626\ldots \;,
\end{equation}
(where $\rme = 2.718281\ldots$ and $\erfc(x)$ denotes an error function \cite{Abramowitz1965a}). Hence $\Delta_{2,2} = 1 - 2\Delta_{2,1} = 0.198748\ldots$. The splittings are very well approximated by the rational sequence $(\frac25,\frac15,\frac25)$. Similarly we get for $k=3$ $\Delta_{3,1}=\Delta_{3,4}=0.349992\ldots$, and $\Delta_{3,2}=\Delta_{3,3}=0.150007\ldots$ leading to the sequence $(\frac{7}{20},\frac{3}{20},\frac{3}{20},\frac{7}{20})$. The plateaus in the integrated density of states and transversal resistivity, respectively, due to the fractional values of $\kappa$, are easily identified in all figures.

\subsection{Resistivity plots}

Early experiments to measure the Hall resistivity were performed with fixed magnetic field while effectively changing the Fermi energy of the system \cite{Klitzing1980a}.
A more common experimental configuration fixes the Fermi energy and records $\rho_{xx}$ and $\rho_{xy}$ as a function of $\mgf $ \cite{Paalanen1982a}.  The dependence is shown in fig.~\ref{fig:HallB}.  In both setups, the transversal current vanishes ($j_y=0$), yielding an implicit expression for the Hall field ${\cal E}_y$:
\begin{equation}
\label{eq:doit}
{\cal E}_y = \rho_{xy}(\mgf ,{\cal E}_y,E_F)\, j_x,
\end{equation}
that can be solved for ${\cal E}_y$ under the constraint of a constant longitudinal current density $j_x$.

The physics behind (\ref{eq:doit}) is that we allow for charge fluctuations due to the presence of external particle reservoirs that feed the electron current. It is interesting to see how the changing number $N(E_F, {\cal B})$ of mobile carriers that participate in the transport generates plateaus in the resistivity. They can be regarded as quantum fluctuations about the classical Hall resistance (see fig.2). Fluctuations in $N(E_F)$ may in turn lead to changes in $E_F$, a possibility which we have not taken into account here. This problem is difficult; it cannot be solved within the simple approach presented in this paper. 

In practice, interpretation of the plots is impeded by the simultaneous presence of two spin populations.  Depending on the spin splitting, we expect different fractions of the resistivity quantum $h/e^2$ to occur in the QHE.  Let us first consider the case of a very small spin splitting ($g\rightarrow 0$).  In this case, the spin dependent (integrated) density of states differs from the spin independent quantities merely by a factor of two.  According to eqs.~(\ref{eq:integerQHE}) and (\ref{eq:non-integerQHE}), the principal plateaus of the Hall resistivity due to a completed Landau level are then located at $\rho_{xy}=2\pi\hbar/(f e^2)$ with $f=2,4,6,\ldots$, while the additional plateaus related to the nodes of $u_k(x)$ are characterized by $f=3,\frac{24}5,\frac{26}5,\frac{67}{10},7$ etc.  The fairly different nature of the two sets of plateaus is illustrated in figs.~\ref{fig:DOS} and \ref{fig:HallB}.
Once the spin degeneracy is lifted ($g\neq 0$), due to the energy splitting $g\hbar\omega_L$ in eq.~(\ref{eq:spin}), the Landau levels observed in the density of states $n_{\uparrow,\downarrow}(E)$ effectively double.  Integer QHE plateaus in $\rho_{xy}$ are then seen for all integer $f$, while the onset of the non-integer effect depends on the value of $g$.  E.~g., for $g=\frac12$ the list of expected features comprises plateaus at $f=1,2,\frac52,3,\frac72,4,\frac{22}5,\frac{23}5,5,\ldots$ which manifest themselves in fig.~\ref{fig:HallB}.

\subsection{Breakdown of the quantum Hall effect}

From eq.~(\ref{eq:LLoverlap}) one concludes that the disappearance of the plateau between two adjacent Landau levels $k-1$, $k$ happens at a critical Hall field ${\cal E}_{\rm crit}$
\begin{equation}
{\cal E}_{\rm crit} = \frac{\sqrt{e\hbar}}m \frac{\mgf ^{3/2}}{\sqrt{2k-1} + \sqrt{2k+1}}.
\end{equation}
Strong experimental evidence for this dependence is reported in
Refs.~\cite{Bliek1986a,Kawaji1993a,Shimada1998a}. With increasing electric field the modulated Landau levels broaden and fill in the gaps in the density of states separating  the integer QHE.  The zeroes of the density of states inside the Landau levels are able to withstand stronger electric fields (see fig.~\ref{fig:DOS}).  This is particularly true for plateaus of half-integer index ($\kappa_{1,1} = \frac32$, $\kappa_{3,2}=\frac72$, etc.):  Their position only marginally depends on the exact value of the Hall field $\elf$, which renders them favourable for experimental detection of the effect.  Finally, if $\elf$ considerably exceeds the critical value ${\cal E}_{\rm crit}$, many Landau levels overlap, and $\rho_{xy} \sim 2\pi\hbar^2\mgf /(emE_F)$ asymptotically becomes a linear function of $\mgf $, thus recovering the classical Hall effect.

\section{Conclusion}

We have analytically calculated how the dynamics of electrons that are injected into crossed electric and magnetic fields, depends on the field parameters and on the energy of the electrons. The quantum mechanical $\ELF\times\MGF$ drift was analyzed in terms of the imaginary part of the appropriate Green function which, in turn, can be related to the density of states for those electrons that participate in the drift. It was demonstrated how the addition of an electric field will lead to important changes in the density of states as compared to those of a purely magnetic field: 
The discrete delta distributions are broadened into smooth functions whose width increases linearly with the applied electric field. Their shape coincides with the density distribution of the various oscillator eigenstates. Hence, the density of states for the $k$th Landau level has $k$ zeroes that lead to additional plateau-like structures in $N(E_F)$.
For a non--interacting Fermi gas of electrons the resulting resistivity plots bear remarkable resemblance to actual quantum Hall effect data. Finally, the model predicts a breakdown of the integer quantum Hall effect at critical electric fields in accordance with experimental observations.

\ack

We appreciate stimulating discussions with A.~Lohr.  Partial financial support through scholarships by the Alexander von Humboldt foundation and the Killam trust (C.~B.) and the Deutsche Forschungsgemeinschaft (project number Kl~315/6-1) is gratefully acknowledged.


\section*{References}

\begin{thebibliography}{10}

\bibitem{Agostini1979a}
P.~Agostini, F.~Fabre, G.~Mainfray, G.~Petite, and N.K. Rahman.
\newblock Free-free transitions following six-photon ionization of {X}enon
  atoms.
\newblock {\em Phys. Rev. Lett.}, 42:1127, 1979.

\bibitem{Corkum1993a}
P.B. Corkum.
\newblock Plasma perspective on strong multiphoton ionization.
\newblock {\em Phys. Rev. Lett.}, 71:1994, 1993.

\bibitem{Becker1995a}
W.~Becker, A.~Lohr, and M.~Kleber.
\newblock Light at the end of the tunnel: two- and three-step models in
  intense-field laser-atom physics.
\newblock {\em Quantum Semiclass. Opt.}, 7:423, 1995.

\bibitem{Blondel1996a}
C.~Blondel, C.~Delsart, and F.~Dulieu.
\newblock The photodetachment microscope.
\newblock {\em Phys. Rev. Lett.}, 77:3755, 1996.

\bibitem{Yukich2003a}
J.N. Yukich, T.~Kramer, and C.~Bracher.
\newblock Negative-ion photodetachment in parallel electric and magnetic
  fields.
\newblock {\em Phys. Rev. {\rm A}}, 2003.
\newblock accepted for publication, e-print:arxiv.org/abs/physics/0304039.

\bibitem{Feynman1965a}
R.P. Feynman and A.R. Hibbs.
\newblock {\em Quantum Mechanics and Path Integrals}.
\newblock McGraw-Hill, New York, 1965.

\bibitem{Grosche1998a}
C.~Grosche and F.~Steiner.
\newblock {\em Handbook of Feynman Path Integrals}, volume 145 of {\em Springer
  Tracts in Modern Physics}.
\newblock Springer, Berlin, 1998.

\bibitem{Kramer2001a}
T.~Kramer, C.~Bracher, and M.~Kleber.
\newblock Four-path interference and uncertainty principle in photodetachment
  microscopy.
\newblock {\em Europhys. Lett.}, 56:471, 2001.

\bibitem{Bracher2003a}
C.~Bracher, T.~Kramer, and M.~Kleber.
\newblock Ballistic matter waves with angular momentum: Exact solutions and
  applications.
\newblock {\em Phys.~Rev.~A}, 67:043601, 2003.

\bibitem{Johnson1983a}
B.R. Johnson, J.O. Hirschfelder, and K.-H. Yang.
\newblock Interaction of atoms, molecules, and ions with constant electric and
  magnetic fields.
\newblock {\em Reviews of Modern Physics}, 55:109, 1983.

\bibitem{Dodonov1975a}
V.V. Dodonov, I.A. Malkin, and V.~Man'ko.
\newblock The {G}reen function of the stationary {S}chr{\"o}dinger equation for
  a particle in a uniform magnetic field.
\newblock {\em Phys. Lett. \rm A}, 51:133, 1975.

\bibitem{Prange1987a}
R.E. Prange and S.M. Girvin, editors.
\newblock {\em The Quantum {H}all Effect}.
\newblock Springer, Berlin, 1987.

\bibitem{Grosso2000a}
G.~Grosso and G.P. Parravicini.
\newblock {\em Solid State Physics}.
\newblock Academic Press, New York, 2000.

\bibitem{Nieto1992a}
L.~M. Nieto.
\newblock Green's function for crossed time-dependent electric and magnetic
  fields. {P}hase-space quantum mechanics approach.
\newblock {\em J.~Math.~Phys.}, 33(10):3402--3409, 1992.

\bibitem{Abramowitz1965a}
M.~Abramowitz and I.A. Stegun.
\newblock {\em Handbook of Mathematical Functions}.
\newblock Dover, New York, 1965.

\bibitem{Kramer2000a}
T.~Kramer.
\newblock {\em Quantum ballistic motion in uniform electric and magnetic
  fields}.
\newblock Diploma thesis, Technische Universit{\"a}t M{\"u}nchen, 2000.
\newblock Unpublished.

\bibitem{Halperin1986a}
B.I. Halperin.
\newblock The quantized {H}all effect.
\newblock {\em Sci. Am.}, 254:40, April 1986.

\bibitem{Paalanen1982a}
M.A. Paalanen, D.C. Tsui, and A.C. Gossard.
\newblock Quantized {H}all effect at low temperatures.
\newblock {\em Phys. Rev. {\rm B}}, 25:5566, 1982.

\bibitem{Klitzing1980a}
K.v. Klitzing, G.~Dorda, and M.~Pepper.
\newblock New method for high-accuracy determination of the fine-structure
  constant based on quantized {H}all resistance.
\newblock {\em Phys. Rev. Lett.}, 45:494, 1980.

\bibitem{Bliek1986a}
L.~Bliek, E.~Braun, G.~Hein, V.~Kose, J.~Niemeyer, G.~Weimann, and W.~Schlapp.
\newblock Critical current density for the dissipationless quantum {H}all
  effect.
\newblock {\em Semicond. Sci. Technol.}, 1:110, 1986.

\bibitem{Kawaji1993a}
S.~Kawaji, K.~Hirakawa, and M.~Nagata.
\newblock Device-width dependence of plateau width in quantum {H}all states.
\newblock {\em Physica B}, 184:17--20, 1993.

\bibitem{Shimada1998a}
T.~Shimada, T.~Okamoto, and S.~Kawaji.
\newblock {H}all electric field-dependent broadening of extended state bands in
  {L}andau levels and breakdown of the quantum {H}all effect.
\newblock {\em Physica B}, 249-251:107--110, 1998.

\end{thebibliography}
\end{document}